\documentclass[prb,preprint]{revtex4}
\usepackage{graphicx}

\begin{document}

\title{An Exploration of the Limits of the Maxwell-Boltzmann Distribution}

\author{Yi-Chi Yvette Wu}
\affiliation{Palo Alto Senior High School,
50 Embarcadero Rd, Palo Alto, CA 94301}

\author{L. H. Ford}
\email{ford@cosmos.phy.tufts.edu}
\affiliation{Institute of Cosmology, Department of Physics and Astronomy \\
Tufts University, Medford, Massachusetts 02155, USA}


\begin{abstract}
Selected aspects of the Maxwell-Boltzmann for molecular speeds are discussed, with special attention
to physical effects of the low speed and high speed limits. We use simple approaches to study several topics
which could be included in introductory courses, but are usually only discussed in more advanced or
specialized courses.
\end{abstract}


\maketitle
\baselineskip=14pt	

\section{Introduction}
\label{sec:intro}

The Maxwell-Boltzmann distribution of molecular speeds is a standard topic in introductory physics
courses. Here we argue that there are several less familiar aspects of the distribution which can be
understood by students in an introductory course, and which illustrate subtle physical effects. Some
of these aspects also illustrate concepts in statistics, especially the probabilities of rare events. We begin
in Sect.~\ref{sec:basics} with a review of the basics of the Maxwell-Boltzmann distribution, viewed as
a probability distribution. Section~\ref{sec:low-end} explores the low speed end of the distribution,
and the probabilities for finding especially slow molecules. This allows us to discuss the limits of a
classical description of an ideal gas and the transition to quantum behavior. The opposite, high speed 
end of the distribution, is taken up in Sect.~\ref{sec:high-end}. We give a simple derivation of the
probability for finding a molecular speed above given large value, which does not require knowledge
of special functions. (Usually, this probability is obtained from an asymptotic form of the error function.)
We use this result to illustrate the rapid decrease in probability at high speeds, and to give a simple
discussion of Jeans escape, the loss of planetary atmospheres due to escape from the planet's
gravitational field. Selected aspects of Brownian motion are discussed in Sect.~\ref{sec:Brown}.

\section{Some Basics of the Maxwell-Boltzmann Distribution}
\label{sec:basics}

In this section, we will quickly review some basic aspects of the Maxwell-Boltzmann distribution,
which are covered in most introductory physics texts. The key assumption is a collection of free
classical particles in thermal equilibrium at temperature $T$. The general form of the Maxwell-Boltzmann 
distribution states that the probability of finding a particle with energy $E$ is proportional to the
Boltzmann factor,
$\exp(-E/kT)$, where $k$ is Boltzmann's constant. If $E = \frac{1}{2} m\, v^2$, the kinetic energy of
a particle on mass $m$ and speed $v$, then the normalized probability distribution becomes
\begin{equation}
P(v) = 4\pi \left(\frac{m}{2\pi k T} \right)^\frac{3}{2} \; v^2 \; \exp\left(-\frac{m v^2}{2 k T}\right) \,.
\label{eq:MB}
\end{equation}
In most textbooks, the Boltzmann factor and hence the distribution are derived using the canonical
ensemble, See Refs.~\onlinecite{MB72,LC07,LSC08} for alternative approaches.
Equation~(\ref{eq:MB}) is the probability distribution for the speed of a single particle, so
\begin{equation}
\int_0^\infty P(v) \, dv =1 \,,
\end{equation}
and $P(v) dv$ is the probability for the particle to have a speed in the interval $(v,v+dv)$.
The maximum value of $P(v)$ occurs at the most probable speed, $v = v_p = \sqrt{2 k T/m}$,
where $P'(v)=0$. The root-mean-squared speed,  defined by 
$v_{\rm rms} = \sqrt{\langle v^2 \rangle}$, is $ v_{\rm rms} = \sqrt{3 k T/m}$. Here
\begin{equation}
\langle v^2 \rangle = \int_0^\infty P(v) \,v^2 \, dv \, .
\end{equation}
For a nitrogen molecule at room temperature, $v_p \approx 420  {\rm m/s}$ and 
$v_{\rm rms}  \approx 520  {\rm m/s}$, so we can write
\begin{equation}
v_p = 420  {\rm m/s}\, \left(\frac{T}{300K}\right)^\frac{1}{2} \,  \left(\frac{m_{N_2}}{m}\right)^\frac{1}{2} \,.
\label{eq:vp}
\end{equation}

\begin{figure}
 \centering
 \includegraphics[scale=0.6]{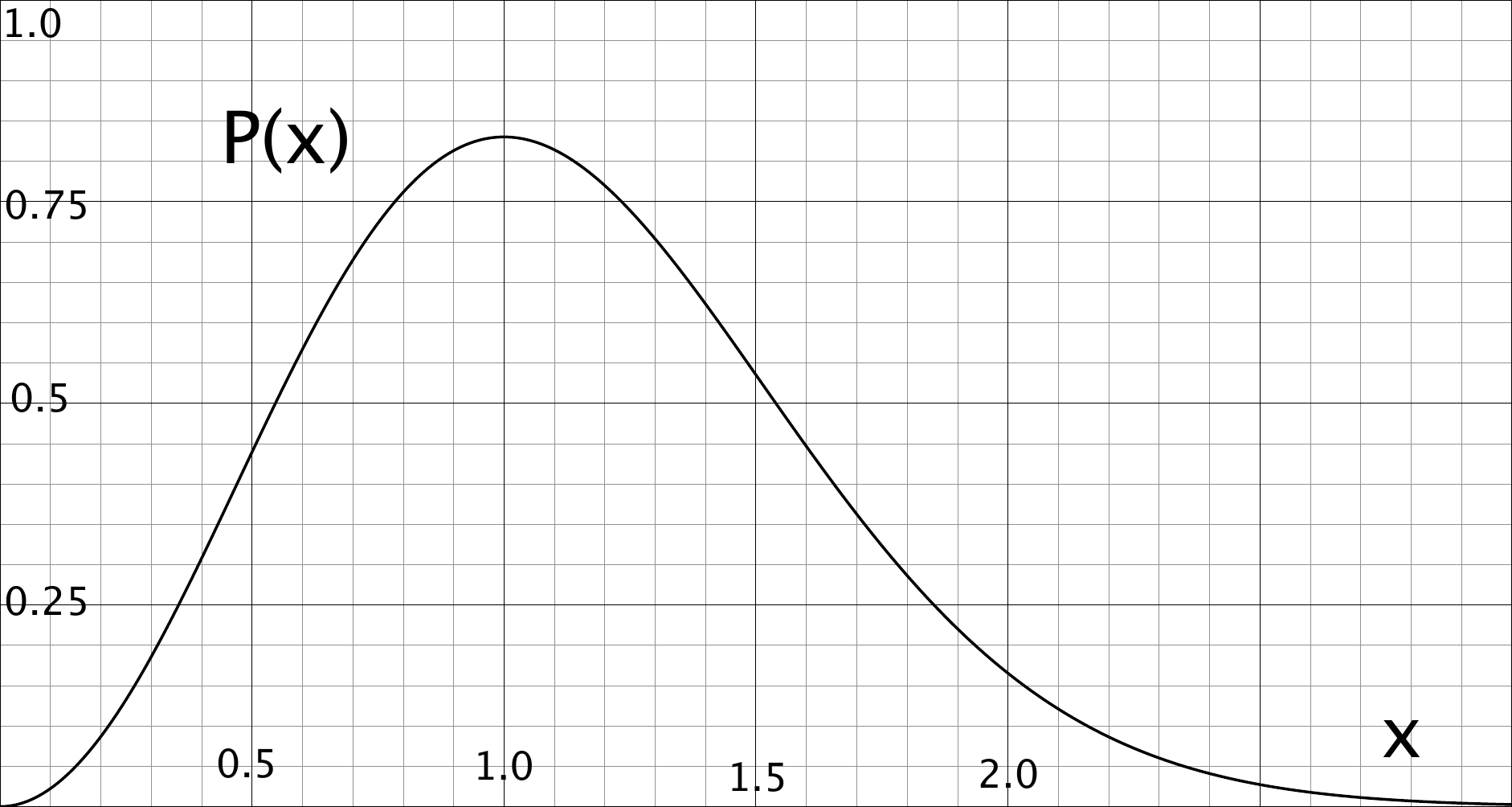}
 \caption{The probability distribution of speeds is given in terms of the dimensionless
 variable $x=v/v_p$, where $v_p$ is the most probable speed.}
 \label{fig:MB}
 \end{figure}

It is convenient to express the probability distribution in terms of the dimensionless
variable $x =v/v_p$, and write
\begin{equation}
P(x) = \frac{4}{\sqrt{\pi}}\, x^2\, {\rm e}^{-x^2} \,,
\label{eq:P}
\end{equation}
where $\int_0^\infty P(x) \, dx =1$. The distribution, $P(x)$, is plotted in Fig.~\ref{fig:MB}.
 
 It is worth trying to understand intuitively why $P(x)$ has a maximum, and falls to zero for both
 high speeds and low speeds. The high speed tail, on the right side of Fig.~\ref{fig:MB} is the
 easier to understand. In thermal equilibrium, with a range in possible particle speeds, there
 will always be a few particles with speeds above the average. However, the ${\rm e}^{-x^2}$
 factor in Eq.~(\ref{eq:P}) strongly suppresses the probability for the speed to be many times
larger than $v_p$. This arises because very fast particles will quickly collide with other particles,
and typically lose energy in the collision. The low speed end of the distribution, where $v \ll v_p$
is often ascribed in advanced texts to ``small phase space volume". However, this is really a way
of saying that it is very unlikely for a particle to sit nearly at rest for a long time when surrounded by 
other rapidly moving particles. The slow particle will be be hit by faster particle and will typically
gain kinetic energy in the collision.

There are some useful quantities which can be estimated from the parameters
of the gas. One of these is the mean free path, which is approximately
\begin{equation}
\ell \approx \frac{1}{\sigma \, \rho} \,,
\label{eq:mfp}
\end{equation}
where $\rho$ is the particle number density and $\sigma$ is the collision cross section.  There is an
average spatial volume of $1/\rho$ for each particle, and a collision becomes likely after the particle
has swept out a volume of $ \sigma\, \ell  \approx 1/\rho$. The typical time between collision is then
$\tau \approx \ell/v_{\rm rms}$. A second useful quantity is the flux of particles striking a surface.
If all particles were moving in the $+x$-direction, taken to be normal to the surface, with speed $v_x$,
then this flux would be $\Phi = n\, v_x$. In the gas in thermal equilibrium, we must have
\begin{equation}
\Phi = c \, \rho\, v_{\rm rms}\,,
\label{eq:flux}
\end{equation}
where $c$ is a numerical constant somewhat less than one. An estimate which imagines that
one-third of the particles are moving along each of the three space directions, and of those moving
along the $x$-axis, one-half are moving in the $+x$-direction, leads to $c =1/6$, An exact calculation
leads to $c=1/4$. See Ref~\onlinecite{Reif} for the details.

\section{Low Velocity Effects}
\label{sec:low-end}
\subsection{Low Velocity End of the Distribution}

Here we will explore some aspects of the low velocity end of the probability distribution, the left side
of Fig.~\ref{fig:MB}, where $x \ll 1$. First, define a cummulative probability by
\begin{equation}
{\cal P}(x < x_0) = \int_0^{x_0} dx\, P(x) \,,
\label{eq:cumm}
\end{equation}
which is the probability of finding that a particle has $x < x_0$, and hence a speed less than $x_0\, v_p$.
If $x_0 \ll 1$, and hence $x \ll 1$, the probability distribution is approximately
\begin{equation}
P(x) \approx \frac{4}{\sqrt{\pi}}\, x^2\, .
\end{equation}
Consequently,
\begin{equation}
{\cal P}(x < x_0) \approx \frac{4}{3\,\sqrt{\pi}}\, {x_0}^3\, .
\label{eq:low-v}
\end{equation}

As discussed in the previous section, the small value of ${\cal P}(x < x_0)$ for small $x_0$ reflects
the improbability that a particle can be nearly at rest at finite temperature. This can be made more
explicit with some examples. The expected speed of the slowest nitrogen molecule in a box at room
temperature and atmospheric pressure with a volume of one cubic meter containing about $10^{25}$ 
molecules, corresponds to 
${\cal P}(x < x_0) \approx 10^{-25}$, leading to the estimate $v_{\rm min} \approx 1\mu {\rm m/s}$.

\subsection{Fraction of ``Quantum Particles"}
\label{sec:quantum}

Here we explore the limits of  a classical description of an ideal gas.
This description fails due to quantum effects when the deBroglie wavelength, 
\begin{equation}
\lambda_{dB} =\frac{h}{m v}
\end{equation}
 of the majority of the particles becomes comparable to the average separation, $d$, of the particles. 
 Here $h$ is Planck's constant. This result arises from 
detailed calculations in quantum statistical mechanics, but can be understood intuitively as follows: 
The classical limit of quantum mechanics occurs when the particles can be described by localized
wavepackets, whose size is small compared to other length scales.  In the case of an ideal gas, this
other scale is $d$. Furthermore, the size of a wavepacket is always larger than $\lambda_{dB}$, so
the classical limit is only possible if $\lambda_{dB} \ll d$.

The ideal gas law,
\begin{equation}
PV =N k T\,,
\label{eq:ideal}
\end{equation}
leads to an expression for $d$,
\begin{equation}
d = \left(\frac{V}{N}\right)^\frac{1}{3} = \left(\frac{k T}{P}\right)^\frac{1}{3} \,.
\end{equation}
This can also be written as
\begin{equation}
d = 3.5 \times 10^{-9} {\rm m}  \left(\frac{T}{300K}\right)^\frac{1}{3}  \left(\frac{1 {\rm atm}}{P}\right)^\frac{1}{3}\,.
\label{eq:d}
\end{equation}
We can estimate the typical deBroglie wavelength of the particles in a gas by setting $v = v_p$,
and writing
\begin{equation}
\lambda_{dB}(v_p) = \frac{h}{m v_p} = \frac{h}{ \sqrt{2 m k T}} \,.
\end{equation} 
This can be written as
\begin{equation}
\lambda_{dB}(v_p) = 8.9 \times 10^{-11} {\rm m}\,\left(\frac{300K}{T}\right)^\frac{1}{2}\,
 \left(\frac{m_{\rm He}}{m}\right)^\frac{1}{2}\,  , 
 \label{eq:lambda-vp}
\end{equation}
where $m_{\rm He}$ is the mass of a helium atom.
The fractional quantum corrections to the classical equation of state, Eq.~(\ref{eq:ideal}) are of 
order~\cite{Huang}
$(\lambda_{dB}(v_p)/d)^3$, and hence are about $10^{-5}$ for helium at room temperature and 
atmospheric pressure, and smaller for heavier molecules, so the classical ideal gas description is 
very good.

However, Eq.~(\ref{eq:low-v}) predicts that there will always be some probability for a particle
to be ``quantum" in the sense that $\lambda_{dB} > d$. We can estimate this probability by letting
\begin{equation}
x_0 = \frac{\lambda_{dB}(v_p)}{d} = 2.6 \times 10^{-2} \,\left(\frac{300K}{T}\right)^\frac{1}{6}\,
 \left(\frac{m_{\rm He}}{m}\right)^\frac{1}{2}\, \left(\frac{P}{1 {\rm atm}}\right)^\frac{1}{3}\,,
\end{equation}
where we have used Eqs.~(\ref{eq:d}) and (\ref{eq:lambda-vp}). This leads to an estimate of the
probability that  $\lambda_{dB} > d$ of
\begin{equation}
{\cal P}(x < x_0) \approx 1.3 \times 10^{-5}   \,\left(\frac{300K}{T}\right)^\frac{1}{2}\,
 \left(\frac{m_{\rm He}}{m}\right)^\frac{3}{2}\, \left(\frac{P}{1 {\rm atm}}\right)\, .
 \label{eq:quant-frac}
\end{equation}
Note that this result is likely to be an underestimate at low temperatures, as helium becomes a superfluid
at $T = 2.17 K$. However by this temperature, a description based on the Maxwell-Boltzmann distribution,
Eq.~(\ref{eq:MB}), needs to be replaced by one based on Bose-Einstein statistics. When the fraction of
quantum particles is small, there is no known effect which they produce, apart from a small correction
to the equation of state. However, we will see in Sect.~\ref{sec:escape}, that sometimes a small fraction
of molecules can have important effects.

\section{High Velocity Effects}
\label{sec:high-end}
\subsection{High Velocity End of the Distribution}

Let us begin by defining an analog of Eq.~(\ref{eq:cumm}) for the upper end of the probability
distribution:
\begin{equation}
{\cal P}(x > x_0) = \int_{x_0}^\infty dx\, P(x) \,.
\label{eq:cumm-h}
\end{equation}
This is the probability of finding $x > x_0$, that is, finding a particle with a speed $v > x_0 \,v_p$.
The integral in Eq.~(\ref{eq:cumm-h}) cannot be performed in terms of elementary functions, but
it does have a simple form when $x_0 \gg 1$. This form can be derived analytically from the asymptotic
form of the error function, but here we give a heuristic derivation.

\begin{figure}
 \centering
 \includegraphics[scale=0.6]{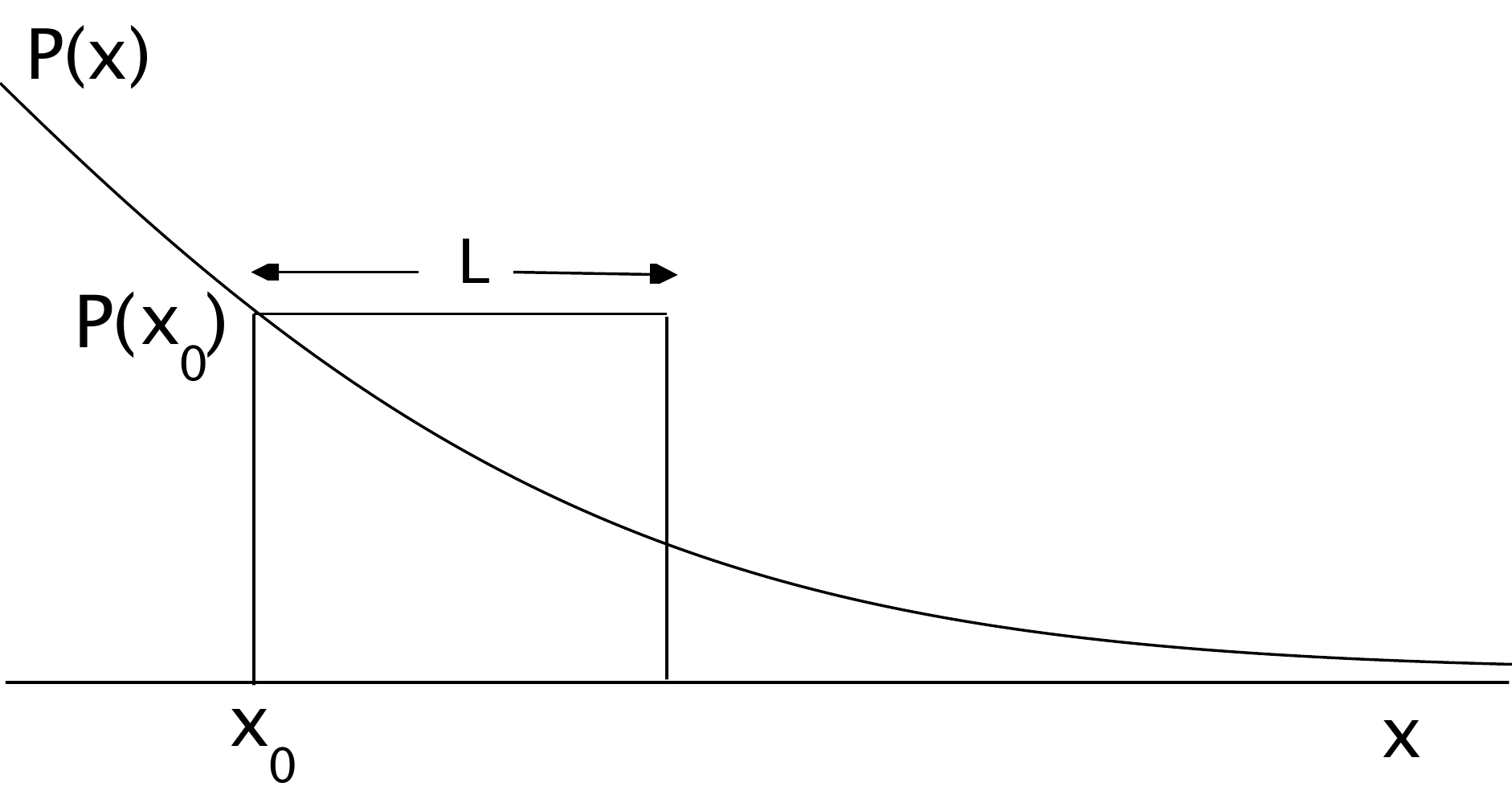}
 \caption{The tail of the probability distribution, $P(x)$, for $x > x_0 \gg 1$ is illustrated. The rectangle of
 height $P(x_0)$ and width $L$ is chosen so that its area,   $P(x_0)\, L$, is equal to the area of the tail,
 $\int_{x_0}^\infty dx\, P(x)$. }
 \label{fig:tail} 
 \end{figure}
The basic idea is illustrated in Fig.~\ref{fig:tail}. The width $L$ of the rectangle is selected so that the
rectangle's area is equal to that under the tail, Eq.~(\ref{eq:cumm-h}).  The trick is to guess a form for $L$
and then confirm it by a numerical experiment. First, we guess that $L$ should decrease with increasing
$x_0$. Given that, the simplest form is $L = a/x_0$, where $a$ is an undetermined constant. The numerical
experiment consists of numerical integration of the right hand side of  Eq.~(\ref{eq:cumm-h}) for various
values of $x_0$, and comparison with
\begin{equation}
{\cal P}(x > x_0) \approx P(x_0)\, L = \frac{a}{x_0}\, P(x_0) \,.
\end{equation}
The result is that $a=1/2$ gives good agreement (within the 1 \% level) between the two forms for $x_0 > 5$.
This confirms our guess for the form of $L$.

This leads to the asymptotic form for $x_0 \gg 1$,
\begin{equation}
{\cal P}(x > x_0)  \approx \frac{2}{\sqrt{\pi}}\, x_0\, {\rm e}^{-x_0^2} \,.
\label{eq:high-v}
\end{equation}
In fact, $x_0$ does not need to be very large before Eq.~(\ref{eq:high-v}) gives a reasonable
estimate. Even for $x_0 =2$, it agrees with the exact result to about 10 \%.
It is of interest to note that the ${\rm e}^{-x_0^2}$ factor causes ${\cal P}(x > x_0)$ to fall extremely rapidly
with increasing $x_0$. For example, ${\cal P}(x > 5) = 7.8 \times 10^{-11}$, meaning that fewer than
one in $10^{-10}$ of the particles has a speed $v > 5\, v_p$, or $v > 4\, v_{\rm rms}$. Nonetheless,
this small fraction of particles can have important effects.

It is important to remember that the Maxwell-Boltzmann distribution applies to a gas in thermal equilibrium.
In the Earth's atmosphere there can be effects, such as collisions with cosmic rays, which could produce
more high speed molecules than predicted by Eq.~(\ref{eq:high-v}). If this happens, there is  a departure
from equilibrium, so the Maxwell-Boltzmann distribution is no longer exactly valid.

\subsection{Escape of Planetary Atmospheres}
\label{sec:escape}

One consequence of the high velocity tail of the distribution can be the escape of gas molecules from
the gravitational field of a planet or moon. If a molecule has a speed greater than the escape velocity,
and can avoid colliding with other molecules, then it can escape into interplanetary space. This
form of escape was first treated by James Jeans~\cite{Jeans}, and is known as ``Jeans escape". The details can
be quite complicated and are discussed in Refs.~\onlinecite{H-S83,CH87,BL04}  , for example. 
In addition, there can be several other
escape mechanisms, such as effects of the solar wind. Here we give a very brief and simplified version.
The escape velocity from a distance $R$ from the center of a planet of mass $M$ is
\begin{equation}
v_{\rm esc} = \sqrt{\frac{2 G M}{R}} \,,
\end{equation}
where $G$ is Newton's constant. For the case of the Earth, $v_{\rm esc} \approx 1.1 \times 10^4 {\rm m/s}$.
Let $x_0 = v_{\rm esc}/v_p$, so Eq.~(\ref{eq:vp}) leads to
\begin{equation}
x_0 \approx 26 \,
\left[ \left(\frac{300 K}{T}\right)\left(\frac{m}{m_{N_2}}\right)\left(\frac{M}{M_E}\right) 
\left(\frac{R_E}{R}\right)\right]^\frac{1}{2} \,,
\label{eq:x0}
\end{equation}
where $M_E$ and $R_E$ are the mass and radius of the Earth, respectively. Equation~(\ref{eq:high-v})
leads to
\begin{equation}
{\cal P}(x > 26) \approx 10^{-292} \,, 
\end{equation}
so there are no nitrogen or oxygen molecules near  the Earth's surface moving faster than the escape
velocity.

Furthermore, a molecule with $v > v_{\rm esc}$ has a reasonable chance to actually escape only if its mean
free path is as large as the distance it needs to travel to leave the planet's atmosphere. This condition
defines the location of the base of the planet's exosphere. From Eq.~(\ref{eq:mfp}),
we see that the mean free path $\ell$ increases as the number density $\rho$  decreases. If we assume that
$\rho$  obeys a Maxwell-Boltzmann distribution, a reasonable guess for the form of $\rho(z)$ is
\begin{equation}
\rho(z) = \rho(0) \,    \exp\left(-\frac{U(z) -U(0)}{k T}\right) \,.
\label{eq:rho}
\end{equation}
Here $z$ is the distance from the surface of the planet, (so $z=0$ on the surface), and $U(z)$ is the 
gravitational potential energy of a molecule as a function of $z$. If $z$ is small compared to the planet's
radius,  we can set $U(0)=0$ and $U(z) \approx m g z$, where $g$ is the gravitational acceleration 
at the surface. In general,
the temperature $T$ is also a function of $z$. We can combine Eqs.~(\ref{eq:mfp}) and (\ref{eq:rho})
to obtain $\ell(z)$, the dependence of the mean free path upon $z$. We can define the height of the base
of the exosphere by $z =z_E$, where
\begin{equation}
z_E = \ell(z_E) \,.
\label{eq:exo}
\end{equation}
This definition is motivated by the notion that once a molecule reaches $z=z_E$ moving outward, it has 
a very good chance to actually escape. By the time it reaches $z= 2z_E$, the mean free path $\ell(z)$
will have increased significantly, so the chances of scattering are still small.

An estimate of flux of escaping particles is given by  Eq.~(\ref{eq:flux}), but with $v_{\rm rms}$ replaced 
by $v_{\rm esc}$, so
\begin{equation}
\Phi{\rm esc} = c \, \rho(z_E)\,{\cal P}(x > x_0)\, v_{\rm esc}\,.
\label{eq:flux-esc}
\end{equation}
Here $\rho(z_E)\,{\cal P}(x > x_0)$ is the number density of molecules at the base of the exosphere with
$v > v_{\rm esc}$.
A more precise value could be obtained by an integration over all $v > v_{\rm esc}$, but $P(v)$ is falling
so rapidly with increasing $v$ that this is unnecessary for an approximate treatment.
The total number of molecules of a given type is given by
\begin{equation}
N = 4 \pi \int_0^\infty (R+z)^2 \, \rho(z) \, dz \,,
\end{equation}
where $\rho(z)$ is the number density of that molecule. The rate of decrease of $N$ due to Jean's escape
is
\begin{equation}
\dot{N} = - A\, \Phi{\rm esc}\,,
\label{eq:dotN}
\end{equation}
where $A$ is the area of the base of the exosphere. The characteristic time for a significant fraction of this 
type of molecule to escape is
\begin{equation}
t_{\rm esc} = \frac{N}{|\dot{N}|} \,.
\label{eq:tesc}
\end{equation}

Here we make a rough estimate of $N$ in terms of $\rho(0)$ and $g$. Let the characteristic height of the
atmosphere, below which a majority of the molecules are located, be $z_0$, defined by 
$\rho(z_0) = {\rm e}^{-1} \, \rho(0)$. If $z_0 \ll R$, then we can write
\begin{equation}
z_0 \approx \frac{k T}{m g}\,.
\end{equation}
For air molecules at about $T =300 K$, this leads to $z_0 \approx 10 {\rm km}$, which is a reasonable
estimate for the Earth. Note that typically $z_0 \ll z_E$.  We can approximate the planet's atmosphere
as a thin shell of area $4\pi R^2$ and thickness $z_0$, and write
\begin{equation}
N \approx 4 \pi\, R^2\, z_0\, \rho(0) \,.
\end{equation}
If $z_E \ll R$, so the area from which molecules escape is also about  $4\pi R^2$, we can combine our
estimate of $N$ with Eqs.~(\ref{eq:flux-esc}),  (\ref{eq:dotN}), and (\ref{eq:tesc}) to give an estimate of the
escape time:
\begin{equation}
t_{\rm esc} \approx \left(\frac{z_0}{v_{\rm esc}}\right)\, \left(\frac{\rho(0) }{\rho(z_E)}\right)\,
 \left(\frac{1}{{\cal P}(x > x_0)}\right)\,.
 \label{eq:tesc2}
\end{equation}
Here we ignore numerical factors of order one, such as $1/c$.
The factor of $z_0/v_{\rm esc}$ represents the time required for a molecule to traverse the lower atmosphere
at the escape velocity if there were no scattering, This time is about 1~second in the case of the Earth.
The other two factors are large dimensionless numbers. The very low density of molecules at the base
of the exosphere compared to the surface of the planet, leads to the factor of ${\rho(0)}/{\rho(z_E)}$, and the 
final factor describes the effect of the fraction of molecules with speeds above the escape velocity.

The height of the base Earth's exosphere can vary, and depends upon conditions such as solar activity,
but is typically a few hundred kilometers. Similarly, the temperature there can also vary, approximately
in the range $500K < T < 1800K$ (See, for example, Fig.~2.7 in Ref.\onlinecite{H-S83}.) 
For a rough approximation, let us assume that
\begin{equation}
{\rho(0)}/{\rho(z_E)} \approx {\rm e}^{z_E/z_0}\,,
\label{eq:rho-ratio} 
\end{equation}
which follows from Eq.~(\ref{eq:rho}) if $T$ is approximately constant. Now all we need to estimate
$t_{\rm esc}$ is the ratio $z_E/z_0$ and the value of $x_0$ from Eq.~(\ref{eq:x0}). For example,
let $z_E/z_0 \approx 20$ and assume that the temperature in Eq.~(\ref{eq:x0}) is $T \approx 1800K$.
For the case of nitrogen molecules, this leads to $x_0 \approx 10.6$, and ${\cal P}(x > x_0) \approx
1.9 \times 10^{-48}$ The resulting escape time is $t_{\rm esc} \approx 8 \times 10^{48} \,{\rm years}$.
The corresponding numbers for hydrogen molecules are $x_0 \approx 2.8$, ${\cal P}(x > x_0) \approx
1.2 \times 10^{-3}$, and $t_{\rm esc} \approx 1.3 \times 10^{4} \,{\rm years}$. This enormous difference
in escape time comes from the lighter mass of $H_2$ leading to a smaller value of $x_0$, and hence
a much greater probability of being above the escape velocity. These estimates are consistent with
the Earth, at an age of about $ 4.6 \times 10^{9} \,{\rm years}$, having nitrogen, but not hydrogen,
persisting in its atmosphere from the time of the Earth's formation.

\section{Brownian Motion}
\label{sec:Brown}

Brownian motion, discovered by the botanist Robert Brown in 1827, is the random motion of small
particles suspended in a fluid, such as water. We now understand this motion as arising
from collisions of the fluid molecules with the particle. This effect can easily be observed in a diluted
homogenized milk in a microscope with a magnification of 400 or more. In this case, the particles are
fat globules with a radius of the order of $0.5\mu {\rm m}$, which appear to be moving randomly at
speeds of a few times $1\mu {\rm m/s}$, due to collisions with water molecules. 

One question which one can ask is whether a given ``jerk" of a fat globule can be due to a collision with
a single water molecule. This question may be answered using the results of Sect~\ref{sec:high-end}.
For a water molecule at room temperature, $v_p \approx 525  {\rm m/s}$, slightly larger than the
corresponding speed of an air molecule due to the lower mass of an $H_2 O$ molecule. We assume
that the water molecule transfers all of its momentum to the globule, so the recoil speed of the globule
is about $v_G = (m_W/m_G) v_W$, where $m_W$ and $m_G$ are the masses of a water molecule
and the globule, respectively, and $v_W$ is the speed of the molecule before the collision. Assume that
the globules are spheres of radius $r$ and density $0.9$ that of water. The required speed of a single
molecule is
\begin{equation}
v_W = 240 v_p\, \left(\frac{r}{1\mu {\rm m}}\right)^3\,   \left(\frac{v_G}{1\mu{\rm m/s}}\right)
         \approx   v_p\, \left(\frac{r}{0.25\mu {\rm m}}\right)^3\,   \left(\frac{v_G}{0.25\mu{\rm m/s}}\right)  \,.     
\end{equation}
This result shows that the answer to this question is extremely sensitive to the size and speed
of the globules. The probability of a single molecule being capable of causing a recoil when $r =1\mu {\rm m}$ 
and $v_G = 1\mu {\rm m/s}$ is ${\cal P}(x > 240) \approx 10^{-25000}$. However, if we reduce both 
the radius and speed of the globule by factors of four, then a significant fraction of the molecules are
capable causing the recoil by themselves. This is another illustration of the powerful effect of the $\exp(-x_0^2)$
term in Eq.~(\ref{eq:high-v}).

Even if a single molecule cannot impart sufficient momentum,  a globule can acquire momentum from the
collective collisions with many molecules. On average, the same number of molecules hit the globule from
either side, so the average momentum acquired is zero. However, this momentum is subject to statistical
fluctuations, so its mean square need not be zero. Here we focus on the average molecules, which have a
speed of about  $v_{\rm rms}\approx 640 {\rm m/s}$ at room temperature. The number of molecules striking a
globule in time $t$ is of order
\begin{equation}
n = \Phi \, A_g\, t \approx \rho \, v_{\rm rms} \, r^2\, t \,,
\end{equation}
where $A_g \approx r^2$ is a measure of the cross sectional area of the globule, and 
$\rho \approx 3.3 \times 10^{28}\, {\rm m}^{-3}$ is the number density of water molecules.
 The momentum carried
by these molecules mostly cancels because of nearly equal numbers of molecules hitting opposite sides
of the globule. However, if the different molecular collisions are uncorrelated with one another, then the
difference in the number of collisions on the right and left sides of the globules will not be exactly equal, but 
will differ by about $\sqrt{n}$. If damping effects could be ignored, this would lead to a net momentum of the 
globule of order
\begin{equation}
p_G \approx \sqrt{n}\, m_W\, v_{\rm rms}\,,
\end{equation}
or an average speed of 
\begin{equation}
v_G \approx r\, \frac{m_W}{m_G} \, \sqrt{\rho\, v_{\rm rms}^3\, t} 
\approx 23{\rm m/s}\, \left(\frac{r}{1\mu {\rm m}}\right)\, \sqrt{\frac{t}{1{\rm s}}}\,.
\label{eq:vG}
\end{equation}

It is clear from the rapid growth of $v_G$ that damping effects cannot be ignored here. Instead, the drag
force described by Stoke's law,
\begin{equation}
F_D = 6\pi \,\eta\, r \, v_G\,,
\end{equation}
becomes important.
Here $\eta \approx 10^{-3} {\rm Pa \,s}$ is the viscosity coefficient for water at room temperature. The
drag force becomes significant when the drag acceleration becomes comparable to the rate of increase
of $v_G$ predicted by Eq,~(\ref{eq:vG}), that is, when
\begin{equation}
\dot{v}_G \approx \frac{F_D}{m_G}\,.
\end{equation} 
This occurs when
\begin{equation}
t \approx 10^{-7}{\rm s}\,  \left(\frac{1\mu {\rm m}}{r}\right)\,,
\end{equation}
and
\begin{equation}
v_G \approx 7.3 \times 10^{-3}{\rm m/s}\, \left(\frac{r}{1\mu {\rm m}}\right)^\frac{1}{2}\,.
\end{equation}

After a very short time, the motion of the globule becomes diffusive and and its displacement
obeys the Einstein relation~\cite{Reif2}
\begin{equation}
\Delta x_{\rm rms} = \sqrt{2\, D \, t}\,,
\end{equation}
where $D$ is the diffusion coefficient given by
\begin{equation}
D = \frac{k T}{6 \pi \eta r} \,.
\end{equation}
This relation describes the motion actually seen in the microscope. For example, with $t= 0.5 {\rm s}$
and $r = {0.5 \mu {\rm m}}$, it predicts
$  v_G = d\Delta x_{\rm rms}/dt \approx 1\mu {\rm m/s}$.

\section{Summary}
\label{sec:sum}

In the previous sections, we have explored selected aspects of the Maxwell-Boltzmann distribution, viewed
as a probability distribution. Particular attention was given to the low speed and high speed limits, which
illustrate the probability of rare events. For example, in Sect.~\ref{sec:quantum}, we examined the probability
of finding molecules moving so slowly that they are no longer classical particles, in the sense that their
deBroglie wavelength exceeds the distance to their nearest neighbor. At room temperature and atmospheric
pressure, this fraction is small, but can exceed $10^{-5}$ for $He$. In Sect.~\ref{sec:high-end}, we looked
at the opposite limit of the especially fast molecules, and found a simple expression for the probability
of such molecules. This result was applied to a simplified discussion of Jeans escape in Sect.~\ref{sec:escape},
whereby those molecules with speeds above the escape velocity can escape into interplanetary space.
Brownian motion was discussed in Sect.~\ref{sec:Brown}. Here we described how this is usually a collective
effect due to the collision of the particle with many molecules, not a few especially rapid molecules. The
statistical fluctuations in momentum transferred to the particle lead to a rapid, short time scale growth
in the mean squared speed of the particle. However, damping effects soon become important on a time
scale much shorter than can be directly observed in a microscope, leading to the diffusive behavior first
described by Einstein.

 \begin{acknowledgments}
This work was supported in part by the National Science Foundation under Grant PHY-1205764.
\end{acknowledgments}

\end{document}